\documentclass[twocolumn,conference]{IEEEtran}
\IEEEoverridecommandlockouts

\usepackage{bm}
\usepackage{color}
\usepackage{graphicx}
\usepackage{epstopdf}
\usepackage{amsmath}
\usepackage{amssymb}
\usepackage{algorithm}
\usepackage{algorithmic}
\usepackage{ulem}
\usepackage{lineno}
\usepackage{lettrine}
\usepackage{subfigure}
\usepackage{multirow}
\usepackage{booktabs}
\usepackage{array}
\usepackage{amsthm}
\usepackage{stfloats}
\usepackage{caption}
\usepackage{setspace}
\usepackage{flushend}
\usepackage{makecell}

\newcommand{\be}{\begin{equation}}
\newcommand{\ee}{\end{equation}}
\newcommand{\bea}{\begin{eqnarray}}
\newcommand{\eea}{\end{eqnarray}}
\newcommand{\ba}{\begin{array}}
\newcommand{\ea}{\end{array}}

\newcommand{\Rmnum}[1]{\expandafter\@slowromancap\romannumeral #1@}
\makeatother
\allowdisplaybreaks[4]
\title{Low-Range-Sidelobe Waveform Design for MIMO-OFDM ISAC Systems}

\author{ \IEEEauthorblockN{Peishi Li$^{\dag}$, Zichao Xiao$^{\dag}$, Ming Li$^{\dag}$, Rang Liu$^{\ddag}$,   and Qian Liu$^{\S}$
		\vspace{-0.0 cm} }
	\IEEEauthorblockA{$^{\dag}$ School of Information and Communication Engineering \\
		Dalian University of Technology, Dalian, Liaoning 116024, China \\
		E-mail: \texttt{ \{lipeishi, xiaozichao\}@mail.dlut.edu.cn, mli@dlut.edu.cn }}
	
	\IEEEauthorblockA{$^{\ddag}$ Center for Pervasive Communications and Computing\\
		University of California, Irvine, CA 92697, USA \\ 
		E-mail: \texttt{rangl2@uci.edu }}
	
	\IEEEauthorblockA{$^{\S}$ School of Computer Science and Technology \\
		Dalian University of Technology, Dalian, Liaoning 116024, China \\ 
		E-mail: \texttt{qianliu@dlut.edu.cn}  \\\; }}

\begin{document}
	\maketitle
	\pagestyle{empty}
	\begin{abstract}
		Integrated sensing and communication (ISAC) is a promising technology in future wireless systems owing to its efficient hardware and spectrum utilization. In this paper, we consider a multi-input multi-output (MIMO) orthogonal frequency division multiplexing (OFDM) ISAC system and propose a novel waveform design to provide better radar ranging performance by taking range sidelobe suppression into consideration. In specific, we aim to design the MIMO-OFDM dual-function waveform to minimize its integrated sidelobe level (ISL) while satisfying the quality of service (QoS) requirements of multi-user communications and the transmit power constraint. To achieve a lower ISL, the symbol-level precoding (SLP) technique is employed to fully exploit the degrees of freedom (DoFs) of the waveform design in both temporal and spatial domains. An efficient algorithm utilizing majorization-minimization (MM) framework is developed to solve the non-convex waveform design problem. Simulation results reveal radar ranging performance improvement and demonstrate the benefits of the proposed SLP-based low-range-sidelobe waveform design in ISAC systems.
	\end{abstract}
	

	\section{Introduction}
	Integrated sensing and communication (ISAC) has gained widespread attention from both industry and academia since it has been deemed as a key technique in future wireless networks and autonomous vehicular systems \cite{IEEE 1}. The crucial challenge in ISAC systems lies in the dual-functional waveform design that should balance the performance requirements of both communication and sensing \cite{Rang Liu}. Multi-input multi-output (MIMO) orthogonal frequency division multiplexing (OFDM) has been widely adopted in various wireless systems, owing to its superior ability of providing high-performance wireless communications. Consequently, the utilization of MIMO-OFDM waveform to achieve both communication and sensing functions is one of the most proposing approaches to deploy ISAC in practical networks. Numerous dual-function MIMO-OFDM waveform designs have been proposed by jointly considering different communication and sensing performance metrics to achieve a satisfactory performance trade-off. Typical sensing performance metrics include the signal-to-interference-plus-noise ratio (SINR) at the radar receiver \cite{sinr}, the mean squared error (MSE) of the beampattern \cite{mse}, the Cram\'{e}r-Rao bound (CRB) \cite{CRB}, etc.
	
	The above-mentioned waveform designs mainly focus on the spatial second-order statistics of the transmitted waveforms, which attempt to improve radar sensing performance while enabling beamforming gains for communications. Ironically, the randomness of communication information causes the ISAC dual-function waveform to inherently exhibit a high range sidelobe level, which may severely degrade the radar ranging performance. However, existing dual-function waveform designs ignore the range sidelobe, which is a crucial factor for radar ranging performance. To mitigate range sidelobes, many deterministic sequences that exhibit good aperiodic auto/cross-correlation properties are introduced for radar waveform designs \cite{sequence}. Therefore, a low-range-sidelobe waveform design is becoming a crucial task for ISAC systems.
	
	Fortunately, the recently-emerged symbol-level precoding (SLP) technique, which fully exploits the degrees of freedom (DoFs) of both spatial and temporal domains, can be utilized for low-range-sidelobe waveform designs in MIMO-OFDM ISAC systems. Particularly, by elaborately designing the waveform in each symbol time-slot, the SLP technique can efficiently suppress the range sidelobes and improve the radar ranging performance. On the other hand, from the perspective of multi-user communications, the SLP technique can utilize the symbol information to convert the harmful multi-user interference (MUI) into constructive interference (CI), thus achieving better communication quality of service (QoS). Therefore, SLP is a promising solution to improve the performance of dual-function waveforms in both communication and sensing functions by fully exploiting the DoFs in both spatial and temporal domains. To the best of our knowledge, SLP-based low-range-sidelobe waveform designs for ISAC systems have not been investigated in the existing literature.

	In this paper, we investigate the SLP-based low-range-sidelobe waveform design for an MIMO-OFDM ISAC system, where a multi-antenna base station (BS) performs both downlink multi-user communication and radar sensing functions. In particular, we optimize the SLP-based MIMO-OFDM waveform to minimize the integrated sidelobe level (ISL) of the transmit waveform to suppress its range sidelobe while satisfying the multi-user communication QoS requirements and the transmit power constraint. We develop an efficient algorithm utilizing the majorization-minimization (MM) framework to solve the resulting non-convex waveform design problem, which exploits the unique signal structures, significantly reducing the computational complexity. Finally, simulation results are provided to reveal radar ranging performance improvement and validate the benefits of the proposed SLP-based low-range-sidelobe waveform design in ISAC systems.
	
	\begin{figure}[!t]
		\centering
		\includegraphics[width = 2.7 in]{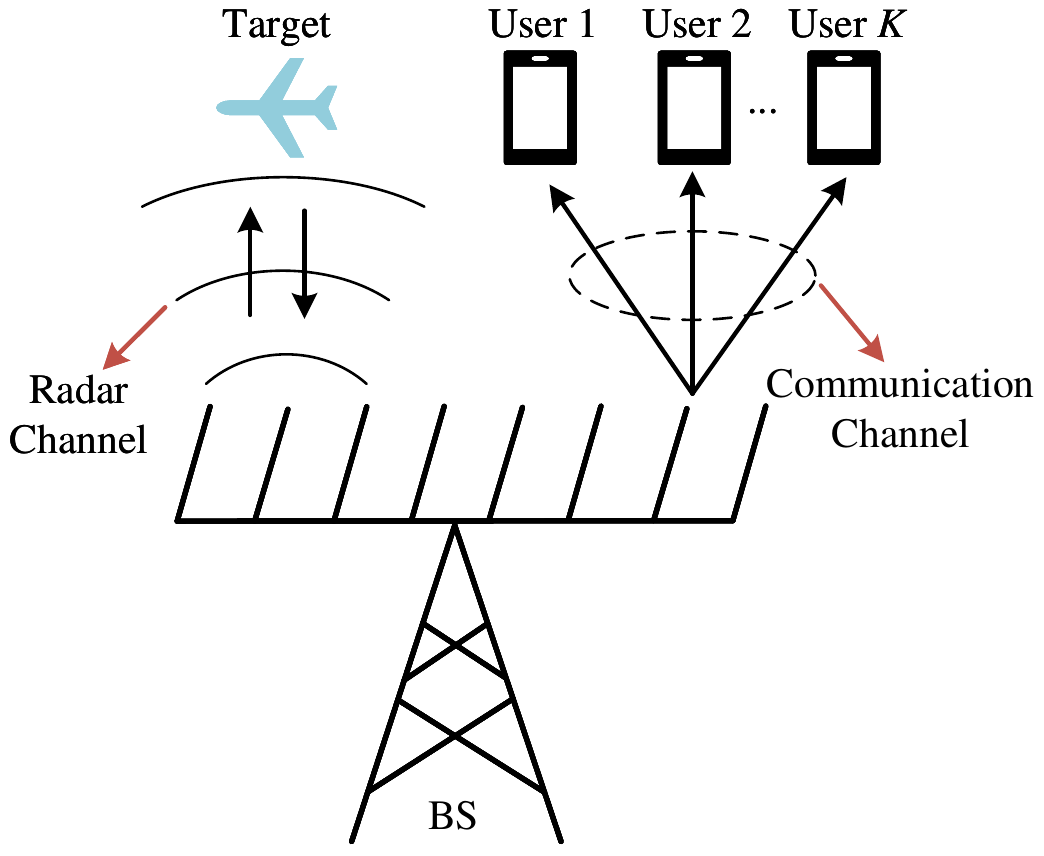}
		\caption{An MIMO-OFDM ISAC system.}
		\label{fig:system_model}\vspace{-3mm}
	\end{figure}
	
	\section{System Model and Problem Formulation}

	\subsection{Transmitted Signal Model}
	
	The considered ISAC system is illustrated in Fig. \ref{fig:system_model}. We assume that the BS is equipped with $N_{\text{t}}$ transmit antennas in a uniform linear array (ULA) and serves $K$ single-antenna communication users while sensing one target with range $R$ in angle $\theta$ using OFDM signals with $N$ subcarriers.
	In specific, the transmitted symbol vector on the $n$-th subcarrier in the $l$-th time-slot is denoted as $\mathbf{s}_{n}[l] \triangleq \left[ s_{n,1}[l], s_{n,2}[l], \dots, s_{n,K}[l] \right]^T \in \mathbb{C}^{K}$.
	The corresponding SLP transmit waveform $\mathbf{x}_{n}[l] \in \mathbb{C}^{N_{\text{t}}}$ is designed to realize functions of delivering information $\mathbf{s}_n[l]$ to $K$ users and sensing the target.
	
	With $\Delta f$ denoting the subcarrier spacing, and $f_{\text{c}}$ the carrier frequency, the transmitted signal is expressed as
	\begin{equation} \label{eq:OFDM transmitted signal}
		\mathbf{x}(t) = \frac{1}{\sqrt{N}} \sum_{n=0}^{N-1} \mathbf{x}_{n}[l] e^{j 2 \pi \left(f_{c}+n\Delta f\right) t},
	\end{equation}
	where $t \in \left[(l-1)(T_{\text{s}}+T_{\text{cp}})+T_{\text{cp}}, l (T_{\text{s}}+T_{\text{cp}}) \right]$, $\forall l$,  $T_{\text{s}}$ is the OFDM symbol duration, and $T_{\text{cp}}$ is the cyclic prefix (CP) duration.
	
	\subsection{Multi-user Communication Model}
	To simplify the description of SLP, we assume that the modulated symbol $s_{n,k}[l]$ is obtained by $\Omega$-phase-shift-keying (PSK) ($\Omega = 2, 4, \dots$). The corresponding SLP transmit waveform $\mathbf{x}_{n}[l]$ is designed to transfer information $s_{n,k}[l]$ to the communication users. The channel impulse response from the BS to the $k$-th user can be modeled by a tapped delay line (TDL) with $U$ taps $\{ \bar{\mathbf{h}}_{1,k}, \dots, \bar{\mathbf{h}}_{u,k} \}$ \cite{tap channel}, where $ \bar{\mathbf{h}}_{u,k} \in \mathbb{C}^{N_{\text{t}}}$ is assumed to be perfectly known at the BS. Then, the received signal on the $n$-th subcarrier at the $k$-th user is given by
	\begin{equation} \label{eq:received signal}
		y_{n, k}[l]=\mathbf{h}_{n, k}^{H} \mathbf{x}_{n}[l] + z_{n, k},
	\end{equation}
	where $\mathbf{h}_{n, k} \in \mathbb{C}^{N_{\text{t}}}$ denotes the frequency-domain channel, and $z_{n, k} \sim \mathcal{CN} (0, \sigma^{2})$ is the additive white Gaussian noise (AWGN).
	
	\begin{figure}[!t]
		\centering
		\includegraphics[width = 178pt]{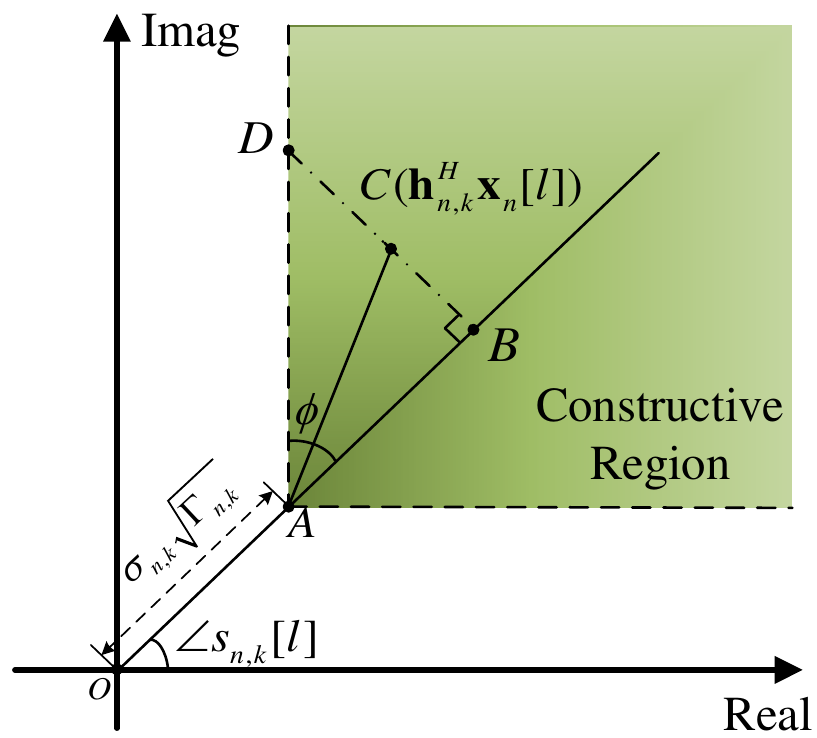}
		\caption{The idea of CI for a QPSK symbol.}
		\label{fig:CR}\vspace{-11pt}
	\end{figure}
	
	SLP converts the harmful MUI into CI by elaborately designing the transmit waveform $\mathbf{x}_{n}[l]$ based on the transmitted symbol $s_{n,k}[l]$ and the channel state information $\mathbf{h}_{n,k}$. To illustrate the concept of CI, we take quadrature-PSK (QPSK) for example, and we assume $s_{n,k}[l] = e^{j\pi/4}$ as the transmit symbol without loss of generality. The idea of CI is shown in Fig. \ref{fig:CR}, where the received noise-free signal is $\mathbf{h}_{n, k}^{H} \mathbf{x}_{n}[l]$, $\phi = \pi / \Omega$, and the decision boundaries for the symbol $e^{j\pi/4}$ are the positive halves of $x$ and $y$ axes. The transmit symbol $s_{n,k}[l]$ can be correctly detected if the received signal $y_{n, k}[l]$ lies in the first quadrant.
	In multi-user systems, the SLP approach designs $\mathbf{x}_n[l]$ to ensure that point $C$ lies in the constructive region, where the MUI is converted into CI that pushes the received signals away from the decision boundaries for improving multi-user communication performance.

	To derive the mathematical formulation for multi-user communication QoS requirements, point $C$ is projected onto the direction of $\overrightarrow{OA}$ at point $B$. We define point $D$ as the intersection of the extension of $\overrightarrow{BC}$ with the boundaries of the constructive region. Then, point $C$ in the constructive region should satisfy $|\overrightarrow{BD}|-|\overrightarrow{BC}| \ge 0$, which can be written as \cite{SLP}
	\begin{equation}\label{eq:communication constraint}
		\begin{aligned}
			&\mathfrak{R}\big\{ \mathbf{h}_{n, k}^{H} \mathbf{x}_{n}[l] e^{-j \angle s_{n,k}[l]} - \sigma \sqrt{\Gamma_{n,k}} \big\}  \tan \phi \\
			& ~~~~~~~~ - \Big|\mathfrak{I}\big\{ \mathbf{h}_{n, k}^{H} \mathbf{x}_{n}[l] e^{-j \angle s_{n,k}[l]} \big\}\Big| \ge 0, ~~ \forall n, k,
		\end{aligned}
	\end{equation}
	where $\Gamma_{n,k}$ is the signal-to-noise ratio (SNR) threshold for user $k$ on the $n$-th subcarrier. Due to space limitations, the detailed derivation is omitted and readers can refer to \cite{SLP}. We utilize some fundamental linear algebra laws to reformulate (\ref{eq:communication constraint}) into an equivalent concise form:
	\begin{equation}\label{eq:comm constraint concise}
		\mathfrak{R}\big\{ \widetilde{\mathbf{h}}_{n, i}^{H} \mathbf{x}_{n}[l]\big\} \ge \gamma_{n,i}, ~~i = 1, 2, \dots, 2K,
	\end{equation}
	where
	\begin{equation}
		\begin{aligned}
			\widetilde{\mathbf{h}}_{n, 2k}^H &\triangleq \mathbf{h}_{n, k}^H e^{-j\angle s_{n, k}}\left(\sin\phi+ e^{-j\frac{\pi}{2}}\cos\phi\right), ~\forall n, k, \\
			\widetilde{\mathbf{h}}_{n, 2k-1}^H &\triangleq \mathbf{h}_{n, k}^H e^{-j\angle s_{n, k}}\left(\sin\phi- e^{-j\frac{\pi}{2}}\cos\phi\right), ~\forall n, k, \\
			\gamma_{n, 2k} &= \gamma_{n, 2k-1} \triangleq \sigma \sqrt{\Gamma_{n, k}} \sin \phi, ~\forall n, k.
		\end{aligned}
	\end{equation}
	In the remainder of this paper, we will employ (\ref{eq:comm constraint concise}) as the multi-user communication performance constraint.

	\subsection{Radar Sensing Model}
	For the stationary target, the radar received echo signal $y_{n}[l]$ can be quantified as \cite{sturm}
	\begin{equation}
		y_n[l] = \beta e^{- j 2 \pi f_{\text{c}} \tau} \mathbf{a}^H(\theta) \mathbf{x}_n[l] e^{-j 2 \pi n \Delta f \tau} + z_{\text{r}},
	\end{equation}
	where $z_{\text{r}}$ is the AWGN at the radar receiver, $\beta \triangleq \sqrt{\frac{\sigma_{\text{RCS}} \lambda^2}{(4\pi)^3 R^4}}$ is the attenuation factor, $\sigma_{\text{RCS}}$ is the target's radar cross section (RCS), $\tau \triangleq 2 R/c_0$ is the round-trip time delay, $c_0$ is the speed of light, and $\mathbf{a}(\theta) \triangleq [ 1, e^{j\frac{2\pi}{\lambda}d\sin(\theta)}, \dots, e^{j\frac{2\pi}{\lambda}(N_{\text{t}}-1)d\sin(\theta)} ]^{T} \in \mathbb{C}^{N_{\text{t}}}$ is the transmit steering vector, with $d$ denoting the antenna spacing and $\lambda$ the wavelength. Generally, range estimation is performed by a correlator bank. Therefore, the radar ranging performance is featured by the auto-correlation of the transmit waveform:
	\begin{equation}\label{eq:continuous range AF}
		r(\tau ) =  \int_{-\infty}^{+\infty} x (t){x^{\ast}}(t - \tau )\text{d}t,
	\end{equation}
	where $x(t) = \mathbf{a}^H (\theta) \mathbf{x}(t)$ is the emitted signal through the $N_{\text{t}}$ antennas. With the CP in OFDM signals, we can approximate the samples of the auto-correlation using the circular auto-correlation \cite{OFDM radar MF}, which can be written as
	\begin{equation} \label{eq:discrete range AF and circular}
		r[m] = \sum_{p = 0}^N x [p] x^{\ast}[\bmod (p - m,N)],
	\end{equation}
	where $m$ is range bin index, $x[p] = x(pT_{\text{c}})$, and $T_{\text{c}}=T_{\text{s}}/N$ is the sampling period.
	
	It can be seen that the auto-correlation function $r[m]$ is random due to the randomness of dual-function waveform. The random communication waveforms may lead to high range sidelobes, rendering poor ranging performance. In this paper, we employ the SLP technique to suppress the range sidelobes by elaborately designing the transmit waveform.
	
	It is widely acknowledged that the auto-correlation with a significant peak at the mainlobe $(m=0)$ and low sidelobes $(m \ne 0)$ offer superior radar ranging performance. However, due to the stochastic nature of the dual-function waveform, the auto-correlation $r[m]$ is a random function. Consequently, the resulting range profile may exhibit a high range sidelobe level, thereby compromising the radar ranging performance. To overcome this drawback, we employ the SLP technique to suppress the range sidelobes by elaborately designing the transmit waveform. The commonly used metric to quantify the sidelobe level is the ISL:
	\begin{equation}\label{eq:range ISL define}
		\xi _r = \sum_{\substack{m = - N+1 \\ m \ne 0}}^{N - 1} |r[m]|^2 = 2\sum\limits_{m = 1}^{N - 1} | r[m]{|^2}.
	\end{equation}
	However, the explicit relationship between ISL and $\mathbf{x}_n[l]$ cannot be found from (\ref{eq:range ISL define}). Using the Parseval equality, a more intuitive analytic expression of ISL can be obtained by \cite{Conflict}
	\begin{equation}\label{eq:R-ISL analytic expression}
		\xi_{r}   =2\Big(\frac{1}{N} \sum_{n=0}^{N-1}\left| \mathbf{a}^{H} \mathbf{x}_{n}  \right|^{4}-\big(\frac{1}{N} \sum_{n=0}^{N-1}\left|\mathbf{a}^{H} \mathbf{x}_{n} \right|^{2}\big)^{2}\Big),
	\end{equation}
	where the time-slot index $[l]$ and angle $\theta$ is omitted for conciseness. Since ISL indicates the self-interference from the sidelobes, the smaller the ISL, the better the radar ranging performance.

	\subsection{Problem Formulation}
	To facilitate the algorithm development, we first convert the ISL (\ref{eq:R-ISL analytic expression}) into an equivalent concise form:
	\begin{equation}\label{eq:R-ISL consice}
		\xi_{r} = \frac{2}{N} \widetilde{\mathbf{x}}^{H} \mathbf{A}^{H} \mathbf{A} \widetilde{\mathbf{x}} - \frac{2}{N^2} (\mathbf{x}^H \widetilde{\mathbf{A}} \mathbf{x})^2,
	\end{equation}
	where we define
	\begin{equation}\label{eq:define}
		\begin{aligned}
			\mathbf{x} & \triangleq \left[ \mathbf{x}_{0}^{T}, \mathbf{x}_{1}^{T}, \cdots, \mathbf{x}_{N-1}^{T} \right]^{T}, ~~~~ \widetilde{\mathbf{x}} \triangleq \text{vec}(\mathbf{x} \mathbf{x}^H), \\
			\mathbf{a}_{n} &\triangleq \mathbf{e}_{n+1} \otimes \mathbf{a}, ~ \mathbf{A}_{n} \triangleq \mathbf{a}_{n} \mathbf{a}_{n}^{H}, ~~\tilde{\mathbf{A}} \triangleq \sum_{n=0}^{N-1} \mathbf{A}_{n}, \\
			\mathbf{A} & \triangleq [ \text{vec}(\mathbf{A}_0), \text{vec}(\mathbf{A}_1), \cdots, \text{vec}(\mathbf{A}_{N-1})]^{H}, \hspace{0.7cm}
		\end{aligned}
	\end{equation}
	and $\mathbf{e}_{n} \in \mathbb{R}^{N}$ indicates the $n$-th column of an $N \times N$ identity matrix. We aim to design the SLP transmit waveform $\mathbf{x}$ to minimize the ISL while satisfying the multi-user communication QoS requirements and the transmit power constraint. Therefore, the optimization problem for SLP-based low-range-sidelobe waveform design can be formulated as
	\begin{subequations}\label{eq:reformulated problem}\begin{align}\label{eq:quartic function}
			&\underset{\mathbf{x}_{n}, \forall n}{\min}~~  \xi_{r} = \frac{2}{N} \widetilde{\mathbf{x}}^{H} \mathbf{A}^{H} \mathbf{A} \widetilde{\mathbf{x}} - \frac{2}{N^2} (\mathbf{x}^H \widetilde{\mathbf{A}} \mathbf{x})^2 \\
			\label{eq:comm constraint}
			&~~\text{s.t.}~~~\mathfrak{R}\big\{ \widetilde{\mathbf{h}}_{n, i}^{H} \mathbf{x}_{n}\big\} \ge \gamma_{n,i},~\forall n,i, \\
			\label{eq:power constraint}
			&~~~~~~~~\sum_{n=0} ^{N-1} \left\| \mathbf{x}_{n} \right\|^{2} \leq P_{\text{0}},
		\end{align}
	\end{subequations}
	where $P_{\text{0}}$ is the transmit power budget. Due to the quartic objective function (\ref{eq:quartic function}), the optimization problem (\ref{eq:reformulated problem}) cannot be easily solved via existing optimization algorithms. Therefore, we will develop an efficient algorithm utilizing the MM framework to solve the waveform design problem.
	
	\vspace{1mm}
	\section{Proposed SLP-based Low-Range-Sidelobe Waveform Design Algorithm}
	
	To solve the resulting non-convex waveform design problem (\ref{eq:reformulated problem}) efficiently, we utilize MM framework \cite{MM} to transform the original problem into a more tractable optimization problem. In specific, we seek an upper-bound surrogate function that locally approximates the objective function (\ref{eq:quartic function}) and minimize the surrogate function in each iteration. The following describes the procedure of deriving the surrogate functions for $\widetilde{\mathbf{x}}^{H} \mathbf{A}^{H} \mathbf{A} \widetilde{\mathbf{x}}$ and $(\mathbf{x}^H \widetilde{\mathbf{A}} \mathbf{x})^2$, respectively.

	\newcounter{TempEqCnt}
	\setcounter{TempEqCnt}{\value{equation}}
	\setcounter{equation}{13}
	\begin{figure*}[!t]
		\begin{subequations}\label{eq:part1 linear}
			\begin{align}
				\widetilde{\mathbf{x}}^{H} \mathbf{A}^{H} \mathbf{A} \widetilde{\mathbf{x}} & \overset{\text{(a)}}{\leq} \lambda_{a} \widetilde{\mathbf{x}}^H \widetilde{\mathbf{x}} + 2 \mathfrak{R} \left\{ \widetilde{\mathbf{x}}^H ( \mathbf{A}^{H} \mathbf{A} -\lambda_a \mathbf{I} ) \widetilde{\mathbf{x}}_t \right\} + \widetilde{\mathbf{x}}_t^H (\lambda_a \mathbf{I} - \mathbf{A}^{H} \mathbf{A}) \widetilde{\mathbf{x}}_t \\
				& \overset{\text{(b)}}{\leq} 2 \mathfrak{R} \left\{ \mathbf{x}^H \mathbf{B} \mathbf{x} \right\} + \lambda_a P_{\text{0}}^2  + \widetilde{\mathbf{x}}_t^H (\lambda_a \mathbf{I} - \mathbf{A}^{H} \mathbf{A}) \widetilde{\mathbf{x}}_t \\
				& \overset{\text{(c)}}{\leq} 2 \mathfrak{R} \left\{ 2 \lambda_b \mathbf{x}^H \mathbf{x} + \mathbf{x}_{t}^H \mathbf{B} \mathbf{x} + \mathbf{x}^H \mathbf{B} \mathbf{x}_{t} - 2 \lambda_b \mathbf{x}^H \mathbf{x}_{t} + \mathbf{x}_{t}^H ( \lambda_b \mathbf{I} - \mathbf{B} ) \mathbf{x}_{t}  \right\} + \lambda_a P_{\text{0}}^2  + \widetilde{\mathbf{x}}_t^H (\lambda_a \mathbf{I} - \mathbf{A}^{H} \mathbf{A}) \widetilde{\mathbf{x}}_t \\
				& \overset{\text{(d)}}{\leq} 2 \mathfrak{R} \left\{ \mathbf{x}^{H} (\mathbf{B} + \mathbf{B}^{H}-2 \lambda_b \mathbf{I}) \mathbf{x}_{t} \right\} + \lambda_a P_{0}^2 + \widetilde{\mathbf{x}}_t^H (\lambda_a \mathbf{I} - \mathbf{A}^{H} \mathbf{A}) \widetilde{\mathbf{x}}_t + 2 \lambda_b P_{\text{0}} + 2 \mathfrak{R} \left\{ \mathbf{x}_{t}^H (\lambda_b \mathbf{I} - \mathbf{B}) \mathbf{x}_{t} \right\}, \\
				& \overset{\text{(d)}}{=} 2 \mathfrak{R} \left\{ \mathbf{x}^{H} (\mathbf{B} + \mathbf{B}^{H}-2 \lambda_b \mathbf{I}) \mathbf{x}_{t} \right\} + c_1,
			\end{align}
		\end{subequations}
		\rule[-0pt]{18.1 cm}{0.05em}
	\end{figure*}
	\setcounter{equation}{\value{TempEqCnt}}

	\vspace{1pt}
	\subsection{Majorizing $\widetilde{\mathbf{x}}^{H} \mathbf{A}^{H} \mathbf{A} \widetilde{\mathbf{x}}$}
	The upper-bound surrogate function of $f(\mathbf{x})$ can be derived according to the second-order Taylor expansion at $\mathbf{x}_t$ as \cite{MM}:
	\setcounter{equation}{14}
	\begin{equation}\label{eq:second-order Taylor}
		f(\mathbf{x}) \! \le \! f(\mathbf{x}_t) + \nabla \!f(\mathbf{x}_t)\!^H (\mathbf{x}-\mathbf{x}_t) +\frac{1}{2} (\mathbf{x}-\mathbf{x}_t)\!^H \mathbf{M} (\mathbf{x}-\mathbf{x}_t),\!
	\end{equation}
	where $\mathbf{M} \succeq \nabla^2 f(\mathbf{x})$. Inspired by (\ref{eq:second-order Taylor}), we can transform $\widetilde{\mathbf{x}}^{H} \mathbf{A}^{H} \mathbf{A} \widetilde{\mathbf{x}}$ into a simple linear function based on the second-order Taylor expansion in steps (a) and (c), respectively. The details of the derivation are presented in (\ref{eq:part1 linear}), where $\mathbf{B} \triangleq \sum_{n=0}^{N-1} \mathbf{A}_{n} \mathbf{x}_{t} \mathbf{x}^H_{t} \mathbf{A}_n^{T} - \lambda_a \mathbf{x}_{t} \mathbf{x}^H_{t}$, $\lambda_a = \lambda_{\text{max}} \{ \mathbf{A}^{H} \mathbf{A} \}$ and $\lambda_b = \lambda_{\text{max}} \{ \mathbf{B} \}$ are the maximum eigenvalues of matrices $\mathbf{A}^{H} \mathbf{A}$ and $\mathbf{B}$, respectively, and the constant term $c_1$ is irrelevant to variable $\mathbf{x}$. With the transmit power constraint (\ref{eq:power constraint}), step (b) is obtained by $	\widetilde{\mathbf{x}}^H \widetilde{\mathbf{x}} \!=\! \text{vec}^H(\mathbf{x}\mathbf{x}^H) \text{vec}(\mathbf{x}\mathbf{x}^H) \! \leq \! P_{\text{0}}^2.$ Similarly, step (d) is derived using $\mathbf{x}^H \mathbf{x} \leq P_{\text{0}}$.

	However, calculating the maximum eigenvalue $\lambda_a$ of the matrix $\mathbf{A}^{H} \mathbf{A}$ with the dimension $N^2 N_{\text{t}}^2 \times N^2 N_{\text{t}}^2$ is computationally challenging. To tackle this issue, we exploit the unique signal structures and find that $\lambda_a$ is a constant that depends only on the number of transmit antennas. Recalling the definition given in (\ref{eq:define}), we can obtain that $\mathbf{A}^H \mathbf{A} = \sum_{n=0}^{N-1} \mathbf{A}_n \otimes \mathbf{A}_n$ and $\mathbf{A}_n^H {\mathbf{A}_{n^{'}}} = \mathbf{0}, \forall n \neq n'$. Since $\mathbf{A}_n$ is a Hermitian matrix, the eigenvectors of $\mathbf{A}_n$ and $\mathbf{A}_{n'}$ are orthogonal. Therefore, we can directly obtain the value of $\lambda_a$ as
	\begin{equation}\label{eq:lambda_derive}
		\begin{aligned}
			\lambda_a   & = \max_{\substack{n = 0, 1, \dots, N-1}} \lambda_{\text{max}} \{ \mathbf{A}_n \otimes \mathbf{A}_n \}   \\
			& = \max_{\substack{n = 0, 1, \dots, N-1}} \lambda_{\text{max}}^2 \{ (\mathbf{e}_{n+1} \otimes \mathbf{a}) (\mathbf{e}_{n+1}^H \otimes \mathbf{a}^H) \} \\
			& = \max_{\substack{n = 0, 1, \dots, N-1}} \lambda_{\text{max}}^2 \{ (\mathbf{e}_{n+1}  \mathbf{e}_{n+1}^H ) \otimes (\mathbf{a} \mathbf{a}^H) \} \\
			& = N_{\text{t}}^2.
		\end{aligned}
	\end{equation}
	
	In this way, we can substantially reduce the computational complexity in majorizing $\widetilde{\mathbf{x}}^{H} \mathbf{A}^{H} \mathbf{A} \widetilde{\mathbf{x}}$. Subsequently, we need to derive the lower-bound surrogate function of $(\mathbf{x}^H \widetilde{\mathbf{A}} \mathbf{x})^2$.

	\subsection{Minorizing $(\mathbf{x}^H \widetilde{\mathbf{A}} \mathbf{x})^2$}
	It is easy to prove that $(\mathbf{x}^H \widetilde{\mathbf{A}} \mathbf{x})^2$ is a convex function with respect to $\mathbf{x}$ by showing that the Hessian matrix $\nabla^2(\mathbf{x}^H \widetilde{\mathbf{A}} \mathbf{x})^2$ is positive semi-definite. Therefore, the lower-bound surrogate function of $(\mathbf{x}^H \widetilde{\mathbf{A}} \mathbf{x})^2$ can be derived according to the first-order Taylor expansion as
	\begin{equation}\label{eq:part2 linear}
		\begin{aligned}
			(\mathbf{x}^H \widetilde{\mathbf{A}} \mathbf{x})^2 & \geq (\mathbf{x}_t^H \widetilde{\mathbf{A}} \mathbf{x}_t)^2 + 4 (\mathbf{x}_t^H \widetilde{\mathbf{A}} \mathbf{x}_t) \mathbf{x}_t^H \widetilde{\mathbf{A}} (\mathbf{x} - \mathbf{x}_t) \\
			&\hspace{58pt}+ 4 (\mathbf{x}_t^H \widetilde{\mathbf{A}} \mathbf{x}_t)  \mathbf{x}_t^T \widetilde{\mathbf{A}}^T (\mathbf{x}^{\ast} - \mathbf{x}_t^{\ast}) \\
			& = 8 \alpha \mathfrak{R} \{ \mathbf{x}^H \widetilde{\mathbf{A}} \mathbf{x}_t \} - 7\alpha^2,
		\end{aligned}
	\end{equation}
	where $\alpha \triangleq \mathbf{x}_t^H \widetilde{\mathbf{A}} \mathbf{x}_t$ for conciseness.
	
	\begin{algorithm}[!t]
		\begin{small}
			\caption{Proposed SLP-based Low-Range-Sidelobe Waveform Design Algorithm}
			\label{alg:1}
			\begin{algorithmic}[1]
				\REQUIRE $\mathbf{h}_{n,k}$, $\Gamma_{n,k}$, $s_{n,k}$, $\forall n,k$, $~\mathbf{a}(\theta)$, $L$, $\phi$, $P_\text{0}$, $\delta_\text{{th}}$.
				\ENSURE $\mathbf{x}^\star$.
				\STATE {Initialize $t:=0$, initialize $\mathbf{x}$.}
				\STATE {Construct $\mathbf{a}_n$, $\mathbf{A}_n$, $\forall n$, $\widetilde{\mathbf{A}}$, $\mathbf{A}$,  $\mathbf{x}$, $\widetilde{\mathbf{x}}$ by (\ref{eq:define}).}
				\STATE {Calculate the objective value $\xi_r$ by (\ref{eq:quartic function}).}
				\STATE {Set $\delta := 1$.}
				\WHILE {$\delta \geq \delta_\text{{th}}$}
				\STATE {$\hat{\xi_r} := \xi_r$.}
				\STATE {Update $\mathbf{b}$ by (\ref{eq:var b}).}
				\STATE {Update $\mathbf{x}_{t+1}$ by solving the problem (\ref{eq:problem after MM}).}
				\STATE {Calculate the objective value $\xi_r$ by (\ref{eq:quartic function}).}
				\STATE {$\delta := \big|(\xi_r - \hat{\xi_r})/ {\xi_r}\big|$.}
				\STATE {$t:=t+1$}
				\ENDWHILE
				\STATE {Return $\mathbf{x}^\star = \mathbf{x}_t$.}
			\end{algorithmic}
		\end{small}
	\end{algorithm}

	Therefore, substituting (\ref{eq:part1 linear}) and (\ref{eq:part2 linear}) into objective function (\ref{eq:quartic function}) leads to the following inequality
	\begin{equation}\label{xi_major}
		\xi_r \le \mathfrak{R} \{ \mathbf{x}^H \mathbf{b} \} + \varepsilon,
	\end{equation}
	where $\varepsilon$ is a constant term and
	\begin{equation} \label{eq:var b}
		\mathbf{b} \triangleq \frac{4}{N}(\mathbf{B}+\mathbf{B}^{H} - 2\lambda_{b} \mathbf{I})\mathbf{x}_{t} - \frac{16\alpha}{N^2}\tilde{\mathbf{A}} \mathbf{x}_{t}.
	\end{equation}
	By ignoring the constant term $\varepsilon$, the optimization problem in each iteration can be formulated as
	\begin{subequations}\label{eq:problem after MM}
		\begin{align}\label{eq:surrogate function}
			&\underset{\mathbf{x}_{n}, \forall n}{\min}~~  \mathfrak{R} \{ \mathbf{x}^H \mathbf{b} \}  \\
			&~~\text{s.t.}~~~\mathfrak{R}\big\{ \widetilde{\mathbf{h}}_{n, i}^{H} \mathbf{x}_{n}\big\} \ge \gamma_{n, i},\forall n,i,\\
			&~~~~~~~~\sum_{n=0} ^{N-1} \left\| \mathbf{x}_{n} \right\|^{2} \leq P_{\text{0}}.
		\end{align}
	\end{subequations}
	It can be seen that the optimization problem (\ref{eq:problem after MM}) is convex, which can be easily solved by various existing convex optimization algorithm.

	With the above derivations, the proposed SLP-based low-range-sidelobe waveform design algorithm is summarized in Algorithm 1, where $\delta_{\text{th}}$ represents the convergence threshold. In summary, the SLP transmit waveform $\mathbf{x}$ is iteratively optimized by solving the problem (\ref{eq:problem after MM}) until convergence.

	\vspace{2mm}	
	\section{Simulation Results}
	
	\begin{figure}[!t]\centering
		\subfigure[Radar range profile for the $K=3$ scenario.]{
			\begin{minipage}[b]{0.45\textwidth}
				\centering
				\includegraphics[width = 3 in]{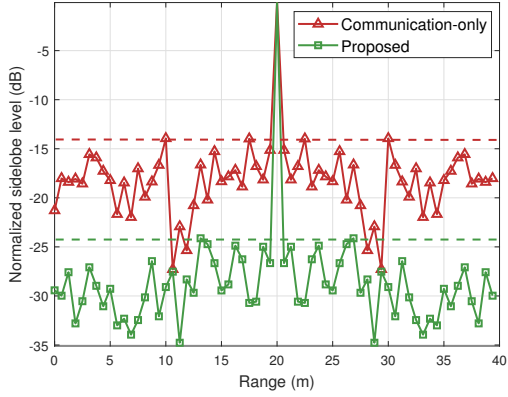}
			\end{minipage}
		}\vspace{1mm}
		\subfigure[Radar range profile for the $K=4$ scenario.]{
			\begin{minipage}[b]{0.45\textwidth}
				\centering
				\includegraphics[width =  3 in]{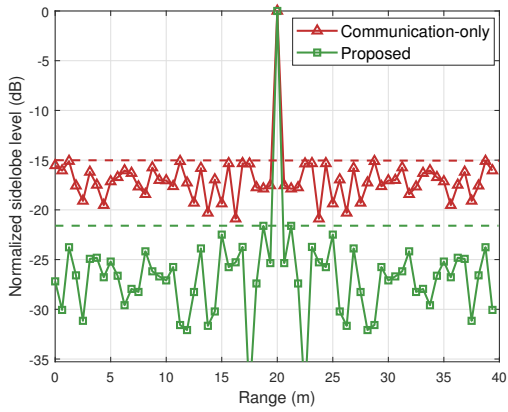}
			\end{minipage}
		}
		\caption{Radar range profiles for a target at 20m, $\Gamma=6$dB, $L = 50$.}
		\label{fig:range profiles}\vspace{-2mm}
	\end{figure}

		In this section, we provide numerical results to evaluate the performance of the proposed SLP-based low-range-sidelobe waveform design for ISAC systems. Without loss of generality, we assume that the BS is equipped with $N_{\text{t}} =8$ antennas with antenna spacing $d = \lambda/2$. The number of subcarriers is $N=64$, the CP length is $N_\text{CP}=N/4$. The communication noise power is $\sigma^2 = 10$dBm, and the QoS requirements are identical for each subcarrier and communication user, i.e., $\Gamma = \Gamma_{n, k}$. In addition, the transmit power is $P_{\text{0}} = 0.5$W and the convergence threshold is set as $\delta_{\text{th}} = 10^{-5}$.
	
	\begin{figure}[!t]\centering
		\subfigure[Communication-only waveform.]{
			\begin{minipage}[b]{0.45\textwidth}
				\centering
				\includegraphics[width = 3 in]{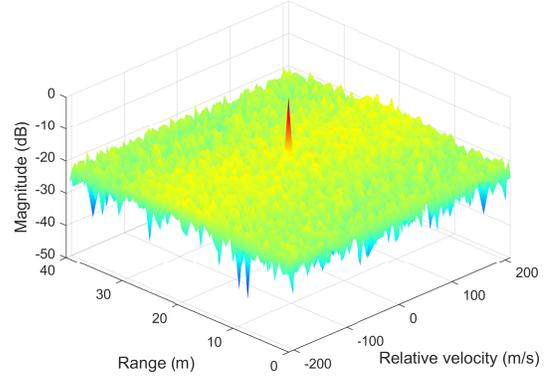}
			\end{minipage}
		}
		\subfigure[Proposed waveform.]{
			\begin{minipage}[b]{0.45\textwidth}
				\centering
				\includegraphics[width =  3 in]{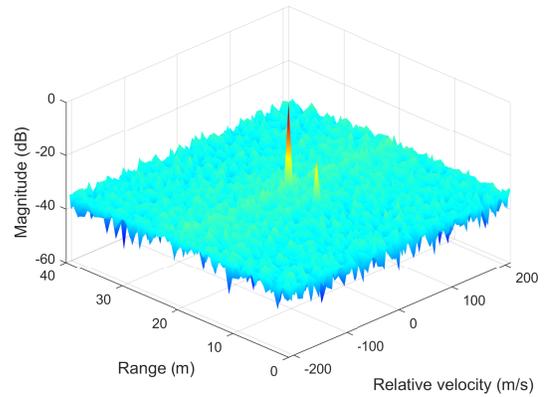}
			\end{minipage}
		}\vspace{-1pt}
		\caption{The range-Doppler map for two stationary targets at ranges 15m and 20m, respectively, $\Gamma=6$dB, $L=256$.}
		\label{fig:RDM}\vspace{-2mm}
	\end{figure}
	
	To validate the performance of the proposed SLP-based low-range-sidelobe waveform design in terms of range sidelobe suppression, we employ the proximate matched filter with circular correlation (PMF-CC) \cite{OFDM radar MF} commonly used in OFDM radar to achieve the range profile. The radar range profiles of different waveforms are shown in Fig. \ref{fig:range profiles}. The communication-only waveform is obtained by minimizing the transmitted power under the communication QoS requirements. It can be seen that the proposed waveform achieves a 10dB range sidelobe reduction for the $K=3$ scenario and a 7dB range sidelobe reduction for the $K=4$ scenario compared with the communication-only waveform, which improves the radar ranging performance while maintaining satisfactory communication QoS.
	
	Next, we demonstrate the radar ranging performance of the proposed SLP-based low-range-sidelobe waveform design in a multi-target scenario. The range-Doppler map is also achieved by PMF-CC in \cite{OFDM radar MF}. In Fig. \ref{fig:RDM}, a strong target with $\sigma_{\text{RCS}}=20$dBsm (e.g. car) and a weak target with $\sigma_{\text{RCS}}=1$dBsm (e.g. pedestrian) are simulated at nearby ranges. For the communication-only waveform, the weak target at 15m is likely to be submerged in the sidelobes of the strong target at 20m, which makes it difficult to be detected. By contrast, for the proposed waveform, the mainlobe level of the weak target is much higher than the sidelobe level so that it can be detected easily. Therefore, the proposed SLP-based low-range-sidelobe waveform design can achieve better radar ranging performance by suppressing the range sidelobes.

\begin{figure}[!t]
		\centering
		\includegraphics[width = 3.2 in]{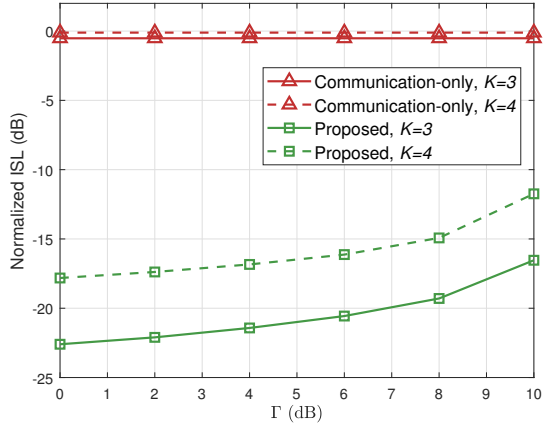}
		\caption{The ISL of different waveforms versus the communication QoS requirements $\Gamma$.}
		\label{fig:ISL_Gamma}\vspace{-2mm}
	\end{figure}

	Furthermore, Fig. \ref{fig:ISL_Gamma} evaluates the ranging performance of different waveforms in terms of the ISL. Compared with the communication-only waveform, the proposed SLP-based low-range-sidelobe waveform design can significantly reduce the ISL of transmit waveform by exploiting more DoFs in the temporal domain, improving the radar ranging performance. Besides, the ISL of the proposed waveform increases with larger $\Gamma$ and $K$, which demonstrates the performance trade-off between multi-user communication and radar sensing.

\begin{figure}[!t]
		\centering
		\includegraphics[width = 3.2in]{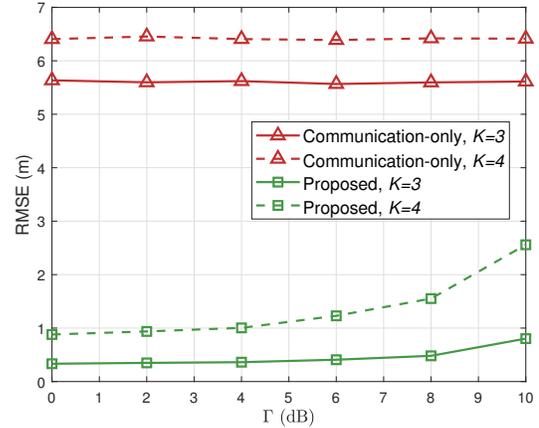}
		\caption{Range estimation RMSE of the weak target versus the communication QoS requirements $\Gamma$.}
		\label{fig:RMSE_Gamma_final}\vspace{-7pt}
	\end{figure}

	Finally, in order to further investigate the impact of range sidelobes on radar ranging performance, we analyze the range estimation root mean square error (RMSE) of the weak target for the case that we attempt to estimate the range of two distinct targets with different RCS. The strong target is located at 20m with $\sigma_{\text{RCS}}=20$dBsm, while the weak target is randomly distributed between 20m to 25m with $\sigma_{\text{RCS}}=1$dBsm. From Fig. \ref{fig:RMSE_Gamma_final}, as the communication QoS requirements $\Gamma$ and the number of communication users $K$ increase, the RMSE of range estimation gradually increases, which further illustrates the trade-off between radar ranging performance and communication QoS. Additionally, our proposed waveform always provides better range estimation performance than the communication-only waveform by suppressing range sidelobes. This demonstrates the superiority of the proposed SLP-based low-range-sidelobe waveform design for range estimation and the effectiveness of utilizing ISL as the sensing performance metric.
	
	\section{Conclusions}
	In this paper, we proposed an SLP-based low-range-sidelobe waveform design for an MIMO-OFDM ISAC system. The ISL of the transmit waveform was minimized while satisfying the multi-user communication QoS requirements and the transmit power constraint. We developed an efficient algorithm utilizing MM framework to solve the waveform design problem. Simulation results were provided to reveal radar ranging performance improvement and demonstrated the benefits of the proposed SLP-based low-range-sidelobe waveform design in ISAC systems.

\end{document}